\documentclass{article}
\usepackage{amsmath,amssymb,amscd,latexsym,amsthm}
\usepackage{graphicx}
\usepackage{slashed}

\author{Peter Woit \\
Department of Mathematics, Columbia University\\
woit@math.columbia.edu}
\title{Towards a Grand Unified Theory of Mathematics and Physics} 
\begin{document}
\maketitle
\begin{abstract}
Wigner's \lq\lq unreasonable effectiveness of mathematics" in physics can be understood as a reflection of a deep and unexpected unity between the fundamental structures of mathematics and of physics.  Some of the history of evidence for this is reviewed, emphasizing developments since Wigner's time and still poorly understood  analogies between number theory and quantum field theory.
\end{abstract}

\section{Introduction}

The question of the relationship of mathematics and physics is as old as the two subjects, and remains very much a mystery.  One point of view is that physics and mathematics are very different subjects, with mathematics providing some useful tools to physicists studying something quite different, while physics provides to mathematics only some interesting examples to study.  In this essay I'll argue that unified theories of fundamental physics are closely linked with some of the great unifying structures that mathematicians have found to underlie much of modern mathematics.  This can be taken as evidence of a possible \lq\lq grand unified theory of physics and mathematics" and motivates the search for a deeper understanding of the known points of contact between the two subjects.

\section{Quantum mechanics and mathematics}

Eugene Wigner's well known essay {\em The Unreasonable Effectiveness of Mathematics in the Natural Sciences} \cite{wigner} (based on a talk delivered in 1959)  concluded with the summary
\begin{quote}
The miracle of the appropriateness of the language of mathematics for the formulation of the laws of physics is a wonderful gift which we neither understand nor deserve.
\end{quote}
While he likely also had in mind Einstein's dramatic success with general relativity and its formulation in terms of the language of differential geometry, he repeatedly invokes examples from quantum mechanics, arguing that these evidenced a miraculous coherence between mathematical formalism and fundamental physics.  

Wigner was himself one of the leading figures in the early application of techniques from group representation theory to the study of atomic spectra, with his 1931 book \cite{wigner-1931} very influential in showing the power of this mathematical language to describe aspects of the surprising and unintuitive new quantum mechanics.  A source for Wigner was mathematician Hermann Weyl's book \cite{weyl} on groups and quantum mechanics, the first edition of which appeared in 1929.  

The language of Lie groups, Lie algebras, and their unitary representations provided a way to simply express the implications for quantum systems of their symmetries, with for instance the behavior of angular momentum operators determined by their role as operators for a unitary representation of the Lie algebra $\mathfrak{so}(3)$ of the rotation group.  The significance of the mathematical language of representation theory went even deeper though than just that of handling symmetries, with the fundamental Heisenberg commutation relations recognized by Weyl as those of a unitary representation of a Lie algebra.  Dirac's remarkable procedure for passing from Hamiltonian mechanics to quantum theory by \lq\lq canonical quantization" is precisely the passage from this Lie algebra to its (unique) irreducible unitary representation (this point of view is developed in detail in \cite{qmbook}).

\section{Quantum mechanics and number theory}

By the time of Wigner's 1959 talk, quantum mechanics and the theory of group representations had developed far beyond the initial insights of the 1920s, with a myriad of close connections between the two subjects.   A few years later connections to quantum mechanics appeared in a very different area of mathematics, with number theorist and algebraic geometer Andr\'e Weil's work \cite{weil1} applying the same Lie group and representation that appears in canonical quantization to questions in number theory.   This involved extending the theory from the real numbers to an arithmetic theory over the rational numbers $\mathbf Q$.  To do this, one thinks of the ring of integers $\mathbf Z$ as a ring of functions on a space called $Spec(\mathbf Z)$, whose points are the prime numbers $p$.  Integers take values at $p$ in the field $\mathbf F_p$ of integers mod $p$. Elements of $\mathbf Q$ can be thought of as rational functions on $Spec(\mathbf Z)$, and at each $p$ one defines a new field called $\mathbf Q_p$ which is the arithmetic analog of Laurent series at  the point $p$.

Weil took the representation of the Heisenberg group constructed by canonical quantization of a classical phase space as just the special case for the field $\mathbf R$ of a more general story that could also be carried through for the local fields $\mathbf Q_p$ and the global field $\mathbf Q$.  He used the recently developed theory of adeles, which packaged the local fields $\mathbf Q_p$ and $\mathbf R$ into a single object, the adele ring $\mathbf A_{\mathbf Q}$, and extended canonical quantization over $\mathbf R$ to this ring.  This formalism then gave a picture which explained a great deal of the theory of theta functions, including results of Carl Ludwig Siegel on generalized theta functions depending on a quadratic form.  For some details of the theory of theta functions from this point of view, see \cite{mumford}.

\section{The Standard Model}

Wigner writes at the end of his essay (referring to the miracle of the effectiveness of mathematics)
\begin{quote}
We should be grateful for it and hope that it will remain valid in future research...
\end{quote}
The next decade or so bore out his hopes well, with the emergence by 1973 of the Standard Model which remains to this day our best theory of fundamental physics.  The Standard Model is defined in terms of  sophisticated geometrical constructs that go well beyond those of general relativity, including the following two central ideas.

\subsection{Yang-Mills gauge theory}
The electromagnetic force can be simply described in terms of a connection on a $U(1)$ principal bundle, the vector potential $A$.  Electric and magnetic fields are components of the curvature of this connection. The action functional is the norm-squared of the curvature, with its Euler-Lagrange equations the Maxwell equations.   The quantization of the theory of just electromagnetic fields (no charged particles) gives the theory of non-interacting photons, and can be performed either in an infinite-dimensional version of canonical quantization or by defining a path integral over the infinite-dimesional space $\mathcal A$ of connections.  Dealing with the infinite-dimensional gauge symmetry of the theory introduces technical difficulties which can be dealt with in several possible ways.

The Standard Model  includes electromagnetism, but also describes the weak and strong forces using the same mathematical language, generalized to the non-abelian case with not just $U(1)$ but also $SU(2)$ and $SU(3)$ bundles.  In yet another demonstration of Wigner's \lq\lq unreasonable effectiveness," the general formalism of connections and curvature on principal bundles developed by geometers during the 1950s turns out to be exactly the right language to formulate this theory.  Just as in the $U(1)$ case, an infinite dimensional group (the gauge group $\mathcal G$) acts on the space of connections $\mathcal A$, preserving the action functional. Handling the implications of this group action is one of the thorniest aspects of the theory.  Unlike the $U(1)$ theory, here even the theory without matter particles is an interacting theory, and much remains to be understood, with a \$1 million Millenium prize for this problem still unclaimed.  A great mystery of the subject remains that of the explanation for this particular set of Lie groups and the relative normalization of the Yang-Mills action terms (why $U(1)\times SU(2)\times SU(3)$?, why the values of the three coupling constants?).  Is there some more fundamental geometrical structure that would explain these choices?

\subsection{Spinor fields and the Dirac operator}

In the Standard Model, matter fields are described by quantizing the infinite dimensional linear space of solutions of a Dirac equation. Such solutions are sections of certain vector bundles, with these vector bundles a tensor product of the spinor bundle over space-time and associated bundles to the principal $U(1)\times SU(2)\times SU(3)$ bundles with connection that describe gauge fields.  This is yet again a construction that is very natural in the formalism of modern geometry, leaving only a small number of unexplained and unmotivated choices (why just the Weyl spinors, and why the intriguing pattern of $U(1)$ hypercharges, together with the fundamental $SU(2)$ and $SU(3)$ representations?)  Again, it seems that we are missing some piece of geometrical structure that would explain these choices.

The field equation here is the Dirac equation, and shortly after Wigner's paper a dramatically new deep connection between mathematics and physics appeared with the discovery by Atiyah and Singer that the Dirac operator plays a fundamental role in their index theorem.  The Atiyah-Singer index theorem is one of the great unifying discoveries of twentieth-century mathematics, bringing together analysis, geometry, and topology in a surprising and non-trivial way.  It soon became clear that K-theory was the right way to formalize some of these ideas, with the Dirac operator playing the role of the fundamental class in K-homology.   For more about K-theory, index theory, and the Dirac operator, the original papers in \cite{atiyah-indextheory} are well-worth reading.

\section{Gauge theory and mathematics}

After 1973, particle physics became somewhat of a victim of its success, as experiment after experiment confirmed precisely the predictions of the Standard Model, culminating with the discovery at the LHC in 2012 of a Higgs particle with the predicted properties.  This period however saw a great deal of progress in the understanding of quantum field theories showing deep and unexpected connections to mathematics.  Beginning with Witten's 1982 paper on {\it{Supersymmetry and Morse theory}} \cite{witten-morse}, the subject of \lq\lq topological quantum field theory" has continued to develop and expand, providing powerful new ideas that have revolutionized several subfields of mathematics.  Among the simplest such theories are quantum mechanical examples with Hamiltonian the square of the Dirac operator, providing a new perspective on the index theorem (see for instance \cite{bgv} and \cite{alvarez-gaume}).   In 1990 Witten was awarded the Fields medal, largely for his 1988 work on Chern-Simons theory \cite{witten-csw}, a three-dimensional quantum field theory that had revolutionary implications for knot theory and three-dimensional topology.  New insights into mathematics and physics continue to come out of the study of this theory in many variants.

These ideas have not been as fruitful in fundamental physics as in mathematics, but there are close connections between the quantum field theories being used and the Standard Model. In particular, Witten's original 1988 TQFT \cite{witten-tqft} (which later led to the Seiberg-Witten equations and major results in four-dimensional topology) is a four-dimensional theory with Yang-Mills and matter fields. It differs from a potentially physical theory only by a subtlety (\lq\lq twisting" of the $N=2$ supersymmetry), while carrying a very simple mathematical interpretation in terms of the de Rham cohomology of a space of connections.   

\section{The Langlands program: number theory and representation theory}

The same period of the late 1960s-early 1970s that saw dramatic progress in particle physics coming from the generalization to non-Abelian gauge fields saw equally dramatic progress in number theory, also involving generalization to non-Abelian groups.   This subject has been described by Edward Frenkel \cite{frenkel} as a \lq\lq Grand Unified Theory of Mathematics":
\begin{quote}
The Langlands Program has emerged in recent years as a blueprint for a Grand Unified Theory of Mathematics. Conceived initially as a bridge between Number Theory and Automorphic Representations, it has now expanded into such areas as Geometry and Quantum Field Theory, weaving together seemingly unrelated disciplines into a web of tantalizing conjectures.
\end{quote}  
In the arithmetic Langlands program one works with objects that have intriguing analogies with the non-abelian gauge groups and spaces of connections of the Standard Model.  Instead of just the adele ring $\mathbf A_{\mathbf Q}$ used by Weil, adele groups such as $SL(2,\mathbf A_{\mathbf Q})$ appear, with different local groups $SL(2,\mathbf Q_p)$ acting at each point $p\in Spec(\mathbf Z)$.  This is tantalizingly like Yang-Mills theory, with its local gauge group acting independently at each point in space-time.  In their 1983 work on the Yang-Mills theory for the case of two-dimensional Riemann surfaces, Atiyah and Bott \cite{atiyah-bott} found intriguing analogies between moduli spaces of connections and the adelic constructions that appear in the Langlands program. 

A fundamental object of the Langlands theory is an \lq\lq automorphic representation," one aspect of which is that of a representation of the adele group (for an introduction to the subject, see \cite{bernstein-gelbart}).  For mathematicians, automorphic representations themselves present fascinating structures to study, with the added interest of detailed conjectures linking them to representations of Galois groups which determine much of the structure of number fields.  In a hazy analogy with Yang-Mills theory, automorphic representations would be analogous to something like representations of the infinite dimensional gauge group, an aspect of the quantum gauge theory formalism that is very poorly understood.  Witten in 1987 \cite{witten-automorphic} had noticed some striking analogies between two-dimensional quantum field theories and automorphic representations, but finally decided not to pursue the subject, commenting recently \cite{witten-ipmu} \lq\lq I concluded reluctantly that the analogy in the form I was developing was way too superficial, so I stopped."

\section{Geometric Langlands}

Mathematicians have for a long time known that there was a detailed analogy between number fields like $\mathbf Q$ and the field of functions on an algebraic curve over a finite field.  The Langlands program applied equally to both contexts, with the second one often technically easier to handle.  There is yet a third subject that fits into this analogy, the subject of Riemann surfaces, thought of as algebraic curves over the field $\mathbf C$, and this was the sort of thing Witten was pursuing in 1987.  The mid-90s saw the development of a much more succesful approach to the subject, which now goes under the name \lq\lq geometric Langlands."   It posits a particular analog of the Langlands program in the context of Riemann surfaces, and has led to a great deal of work bringing together different areas of mathematics, and motivating Frenkel's \lq\lq Grand Unification" characterization.  This remains a very active field of mathematical research, with some signs that ideas developed in it may feed back into the continuing research on the arithmetic case.

A 2004 workshop at the Institute for Advanced Study brought together physicists and mathematicians working on the Langlands program.  It inspired Witten to find a way to relate one of the 4d supersymmetric Yang-Mills theories studied in TQFT to the geometric Langlands context, with details worked out in a paper of Kapustin and Witten \cite{kapustin-witten}. This led to a large number of results linking geometric Langlands and quantum field theories, and work on this topic continues to this day.

\section{Some speculation}

Trying to look into the future, it is difficult to impossible to predict where successful new ideas bringing together quantum field theory and mathematics will come from, but some possibilities are
\begin{itemize}
\item
Our current understanding of quantum field theory is dominated by the idea of defining the theory by starting with a Lagrangian functional, then applying canonical quantization or path integral methods.  It appears however that many interesting QFTs cannot be defined this way, with one example the six-dimensional superconformal $N=(2,0)$ theory that is the focus of much current research.  It seems possible that new definitions of quantum field theory will be found, ones more directly encoding new aspects of the role of representation theory in the subject.
\item
Homological methods have played a large role in modern developments in both mathematics and quantum field theory.  In recent years representation theory has seen the introduction of some new homological methods (\lq\lq Dirac cohomology", for details see \cite{huang-pandzic} and \cite{meinrenken}), with an algebraic analog of the Dirac operator playing a central role.  This may somehow lead to better understanding of the role of the physical Dirac operator, making clear its representation-theoretic significance.
\item
Better understanding of the relationships between geometric Langlands and quantum field theory may arise, of a sort that would allow a unified picture of the arithmetic case, the function field case and the geometric case.   While $Spec(\mathbf Z)$ is often thought of as analogous to a two-real dimensional space, it also can be thought of instead as analogous to a three dimensional one (with primes analogous to knots, see \cite{morishita}), and perhaps it is three-dimensional quantum field theories related to Chern-Simons theory for which new insight into analogies with number theory will be found.
\end{itemize}

Much speculation about new ideas in fundamental physics involves radical changes in our current theory, with the idea that the Standard Model is just a low-energy effective theory approximating something quite different.   While the modern understanding of renormalization allows this possibility,  it does not require it, and one remarkable aspect of the Standard Model is that it is consistent all the way up to much higher energies than we have any hope of probing experimentally.   One can take the theory's consistency with all current data as evidence that the Standard Model may be something rather close to a final theory.  If this is true, the goal is not to overthrow the Standard Model, but to deepen our understanding of it, perhaps through reformulation that will unearth new aspects of its connection to mathematics, especially to representation theory.  Such deeper understanding may provide answers to questions the theory currently cannot address.  In some sense the suggestion here is much more conservative and less radical than much of current theoretical work.  The past history of theoretical particle physics has been dominated by the discovery of new symmetries and ways to exploit them, so the conjecture that this will continue is unexceptional.  The Langlands program and other connected areas of mathematics are based on a wealth of ideas about symmetries, with intriguing relations to quantum field theory, so the hope that progress may come from that direction is not an unreasonable one.

\section{Conclusion}

The possible existence of a conventional $SO(10)$ grand unified theory of particle physics would have no impact on virtually all of the rest of physics and the many different ways in which physicists understand and explore the physical world.  It is equally true that the existence suggested here of a grand unified theory of mathematics and physics would in no way affect most of what mathematicians do as they explore the wide range of known interesting mathematical structures.  Such a theory by itself would not say anything about most of the rich phenomena that mathematicians study, but would just indicate the existence of a distinguished point in the space of all mathematical structures.  The nature of this point might however shed light on some new connections between such structures.  The lesson drawn here from history is that the fundamental laws of physics point not to some randomly chosen mathematical structure, but to an exceptionally special one,  requiring a deep understanding of the mathematical world in order to fully appreciate it.

New understanding of relations between mathematical ideas and ideas in fundamental physics can lead to progress in either field.  From the earliest days of quantum mechanics, representation theory played a major role in the elucidation of atomic spectra and many other physical questions.  Ideas from physics have had a huge impact on mathematics, with the Chern-Simons and Seiberg-Witten theories a recent example of this.  Going forward, information is equally likely to flow from mathematics to physics, or physics to mathematics. 

The huge success of the Standard Model has put particle theory in a difficult position, with little in the way of experimental hints of the right direction to look for a better theory.  In the past history of the subject, progress has almost always come from such experimental hints, but there always has been an alternative way forward, that of pursuing connections to mathematics as a very different sort of guidance.  One can argue that Einstein's successful development of general relativity was an example of this. Little help came from experiment, but a great deal from mathematicians and the powerful new formalism of Riemannian geometry.

During the half-century since Wigner's talk,  the connections between quantum field theory and mathematics that have come to light go far beyond anything that Wigner could have even dreamed of.  Mathematics research continues to progress at a rapid pace, with the last 20 years seeing proofs of some of the most well-known long-standing conjectures (Fermat's Last Theorem and the Poincar\'e Conjecture).   There seems no reason to believe that further insight into the deep connections between the intertwined subjects of fundamental physics and fundamental mathematics will not be found.  Wigner's \lq\lq unreasonable effectiveness" miracle is ultimately a claim that a unity of mathematics and physics exists despite our lack of any good reason to expect it.  We may not deserve to be part of this miracle, but we can and should continue to try and understand it.

\end{document}